\documentclass[12pt]{article}
\usepackage{amsmath,latexsym,amssymb,epsfig,psfrag}
\usepackage{graphicx}  

\def\ltsim{\lower3pt\hbox{$\, \buildrel < \over \sim \, $}}  
\def\gtsim{\lower3pt\hbox{$\, \buildrel > \over \sim \, $}}  
\newcommand{\be}{\begin{equation}}  
\newcommand{\ee}{\end{equation}}  
\def\ga{\mathrel{\raise.3ex\hbox{$>$\kern-.75em\lower1ex\hbox{$\sim$}}}}  
\def\la{\mathrel{\raise.3ex\hbox{$<$\kern-.75em\lower1ex\hbox{$\sim$}}}}

\makeatletter   
\@addtoreset{equation}{section}   
\makeatother

\def\simlt{\stackrel{<}{{}_\sim}}
\def\simgt{\stackrel{>}{{}_\sim}}  

\begin{document}
\baselineskip=16pt   

\begin{titlepage}  

\vskip 2cm
%\begin{flushright}
%{\bf UAB-FT-nnn}
%\end{flushright}
\begin{center}  
\vspace{0.5cm} \Large {\sc 
The MSSM from Scherk-Schwarz\\ Supersymmetry Breaking}
\vspace*{5mm}  
\normalsize  
  
{\bf 
D.~Diego~\footnote{diego@ifae.es},
G.v.~Gersdorff~\footnote{gero@pha.jhu.edu} 
and
M.~Quir\'os~\footnote{quiros@ifae.es} 
}   

\smallskip   
\medskip   
\it{$^{1,\,3}$~Theoretical Physics Group, IFAE}\\ 
\it{E-08193 Bellaterra (Barcelona), Spain}

\smallskip   
\medskip   
\it{$^{2}$~Dept. of Physics and Astronomy}\\ 
\it{Johns Hopkins University, Baltimore, MD 21218}

\smallskip     
\medskip
\it{$^3$~Instituci\'o Catalana de Recerca i Estudis Avan\c{c}ats (ICREA)}

\vskip0.6in 
\end{center}  
   
\centerline{\large\bf Abstract}  

\noindent
We present a five-dimensional model compactified on an interval where
supersymmetry is broken by the Scherk-Schwarz mechanism. The gauge
sector propagates in the bulk, two Higgs hypermultiplets are
quasilocalized, and quark and lepton multiplets localized, in one of
the boundaries. The effective four-dimensional theory is the MSSM with
very heavy gauginos, heavy squarks and light sleptons and
Higgsinos. The soft tree-level squared masses of the Higgs sector can
be negative and they can (partially) cancel the positive one-loop
contributions from the gauge sector. Electroweak symmetry breaking can
then comfortably be triggered by two-loop radiative corrections from
the top-stop sector. The fine tuning required to obtain the
electroweak scale is found to be much smaller than in the MSSM, with
essentially no fine-tuning for few TeV gaugino masses. All bounds from
direct Higgs searches at LEP and from electroweak precision
observables can be satisfied. The lightest supersymmetric particle is
a (Higgsino-like) neutralino that can accomodate the abundance of Dark
Matter consistently with recent WMAP observations.

\vspace*{2mm}   
%\smallskip\newline  
  
\end{titlepage}  
  
\section{\sc Introduction}
\label{introduction}

Experiments are about to probe the physics which is responsible for
electroweak symmetry breaking (EWSB), and to hopefully shed some light
on one of the biggest open questions which arise within the Standard
Model: the origin and size of the electroweak scale.  Supersymmetry
(SUSY) has long been a very promising candidate to provide a
satisfactory explanation of EWSB, as loops of the top-stop sector
naturally drive the Higgs squared mass to negative values, enforcing
the Higgs to acquire a nontrivial vacuum expectation value (VEV).
Moreover these quantum corrections are cutoff at the mass of the stop
itself, thereby explaining the smallness of the electroweak scale as
long as the stop mass does not exceed several TeV. In fact
understanding the mechanism that triggers supersymmetry breaking is
one of the main issues in supersymmetric theories and it should
determine the phenomenology of supersymmetric particles at future
high-energy colliders as the LHC.

On the other hand the existence of extra dimensions is a general
prediction of fundamental (string) theories that aim to unify all
interactions, including gravity, and provide a consistent quantum
description of them. In particular if the radii of the extra
dimensions are as large as the 1/TeV scale~\cite{Antoniadis:1990ew},
matter can propagate in the bulk and the very existence of extra
dimensions can provide new mechanisms for supersymmetry and
electroweak breaking~\cite{Quiros:2003gg}.  It has also been pointed
out that extra dimensions can help to suppress the dangerous flavour
violating interactions of SUSY breaking, as long as quark and lepton
matter is localized on a supersymmetry preserving
brane~\cite{Randall:1998uk}, as we will assume in this paper. Finally
this type of models where matter fields are localized on 3-branes
while the gauge sector propagates in the bulk of extra dimensions
generically appears in intersecting brane
constructions~\cite{Uranga:2005wn}.

An attractive way of breaking supersymmetry (genuine to theories with
extra dimensions) is the Scherk-Schwarz (SS)
mechanism~\cite{Scherk:1978ta} that makes use of twisted boundary
conditions (BC's). In five and six-dimensional supersymmetric theories
the $SU(2)_R$ invariance of the supersymmetry algebra can be used to
break supersymmetry and hence its breaking should primarily be felt by
$SU(2)_R$ doublets (gauginos in vector multiplets and Higgs bosons in
hypermultiplets).  For definiteness we will consider a
five-dimensional (5D) model where the fifth dimension is compactified
on an interval of length $\pi R$ (compactification scale $M_c\equiv
1/R$)~\footnote{We will often use, unless explicitly stated, units
where $R=1$.}.  States propagating in the bulk of the extra dimension
break supersymmetry due to their BC's at the endpoints of the
interval. While each of the BC's preserves an $N=1$ subgroup of the 5D
$N=2$ SUSY, they need not coincide and hence SUSY can be broken
nonlocally. This particular way of breaking SUSY forbids any explicit
local soft breaking terms in the 5D action. This improves the UV
sensitivity, as quantum corrections are cutoff at the scale $1/R$ due
to non-locality~\cite{Delgado:2001ex}.  Furthermore superfields
localized on one of the boundaries only feel supersymmetry breaking at
the loop level.

In this paper we will consider the natural scenario where gauge
multiplets propagate in the bulk of the extra dimension (which is
assumed to be compactified at or above the TeV scale), while quark and
lepton superfields are localized towards one of the
boundaries~\cite{PQ}. In this way one can obtain a reliable
superpartner spectrum where gauginos get tree level masses of the
order the compactification scale, while squark and slepton masses are
radiatively generated. Furthermore all (5D) massless fields are
flavour blind and dangerous flavour nondiagonal interactions are
mediated only by fields with (5D) masses of the order of the cutoff,
$\Lambda$, and hence they are suppressed as $\exp(-\Lambda \pi
R)$~\cite{Randall:1998uk}.

Since the top-quark is localized, the stop mass is generated at
one-loop and EWSB should be triggered at two-loop. A detailed
discussion of this phenomenon can be found in
Refs.~\cite{Barbieri:2002uk} where it was shown that the one-loop
positive contribution to the squared Higgs mass from the gauge sector
cannot be canceled by the two-loop negative contribution from the
top-stop sector. It was concluded that in SM-like models with only one
light Higgs and all quark and lepton fields localized in one of the
boundaries EWSB does not take place (within the uncertainties of the
two-loop calculation) . A possible way out to this problem would be to
somewhat delocalize the left-handed and/or the right-handed top quark
multiplet~\cite{Barbieri:2002uk,Barbieri:2000vh,Delgado:2001si,Marti}. In
this case the corresponding scalar quarks feel SS supersymmetry
breaking at the tree-level and so EWSB proceeds at one-loop. In these
models the degree of delocalization/quasilocalization of fields $\phi$
is controled by the bulk masses $M_\phi$. Then depending on these
masses FCNC~\cite{Delgado:1999sv} are possible by the tree level
exchange of Kaluza-Klein (KK) modes of the gauge bosons and a careful
choice of masses has to be done to avoid them.  Furthermore as it was
pointed out in Ref.~\cite{Ghilencea:2001bw}, in models with only one
Higgs hypermultiplet propagating in the bulk, quadratically divergent
Fayet-Iliopoulos (FI) terms are generated at one loop. Although
consistent with both gauge symmetry and supersymmetry, these terms
introduce a quadratic sensitivity of the Higgs mass to the UV cutoff.
Finally even in models without quadratically divergent FI terms, in
the presence of bulk masses for hypermultiplets linearly divergent FI
terms $\sim M_\phi \Lambda$ may be generated unless special conditions
on the mass matrix are met.

In this article we propose a different solution to the EWSB problem in
models where all quark and lepton superfields are localized on one of
the boundaries and the Higgs hypermultiplets have a bulk mass $M\simgt
M_c$.  As it is well known~\cite{Georgi:2000wb} this can lead to a
localization of the wavefunction of the lightest mode towards one of
the branes (the lightest mode thus becomes quasilocalized) while all
the higher modes typically become very heavy and
decouple~\footnote{For an application of this quasilocalization effect
to matter fields and flavor physics
see~\cite{Marti,Hebecker:2002re,Abe:2004tq}.}. Moreover one can set up
a well defined expansion in powers of $\epsilon=\exp(-M\pi R)$ which
can be carried out to arbitrary high orders. As the strictly localized
limit (where no tree level soft terms can appear) corresponds to
$\mathcal O(\epsilon^0)$ one expects some tree level soft masses in
the Higgs sector of the order $M\epsilon$. Due to the exponential
dependence these can be naturally of the same order as bulk loop
corrections if the compactification scale is taken to be in the TeV
region.  EWSB in this model is favored by two facts:
\begin{itemize} 
\item
For a region of the SS parameter space the tree-level soft masses can
be tachyonic~\cite{Barbieri:2002uk,Diego:2005mu} and can then totally
or partially compensate for the positive contribution from gauge
one-loop radiative corrections. Under these circumstances EWSB will
proceed in a fairly natural fashion triggered by the two-loop
radiative corrections from the top-stop sector.
\item
In a model with two Higgses the condition for EWSB does not
necessarily imply that one of the Higgs masses becomes negative. In
some cases even if the (negative) two-loop correction is not able to
overcome the positive tree-level and one-loop contributions to the
soft masses, EWSB can proceed.
\end{itemize}

Our model contains two light Higgs doublets, as the MSSM, and so
neither FI quadratic nor linear divergences will be generated by
radiative corrections. The low energy theory is the MSSM with a
peculiar spectrum of supersymmetric particles generated by the SS
supersymmetry breaking. At the tree-level the corresponding
supergravity theory would be a no-scale model~\cite{Ellis:1983sf} and
thus no anomaly mediated supersymmetry breaking \cite{Randall:1998uk} will
appear~\cite{Luty:2002hj,dudas}. Moreover due to the smallness of the
SUSY breaking scale and the extreme softness of the SS breaking the
usual fine-tunings of the MSSM can be avoided or at least, to a large
extent, alleviated. Finally the lightest supersymmetric particle (LSP)
of the model is a neutralino that is a good candidate to Dark
Matter. To the best of our knowledge this is the first time that the
MSSM with EWSB is obtained from an extra dimensional model with SS
supersymmetry breaking and all matter localized in a boundary. Note
that SS breaking clearly distinguishes our model from those with
similar bulk field content but with localized SUSY breaking proceeding
at the distant brane, as for instance in
Refs.~\cite{Mirabelli:1997aj,gauginomediation}. In particular, in the
context of gaugino mediation~\cite{gauginomediation}, the
compactification scale is generally very high (grand unification
scale), while in our case it will turn out to be a few TeV. Although
we are giving up MSSM-like high-scale unification, power law running
of gauge couplings make unification at a much lower scale
possible~\cite{Dienes:1998vh}.  Furthermore, based on earlier
work~\cite{Seiberg:1996bd} it has been pointed out
in~\cite{Hebecker:2002vm} that this running should rather be
interpreted as power-law threshold corrections which are exactly
calculable due to the bulk $N=2$ supersymmetry, thus opening up the
fascinating possibility to construct extremely predictive models of
grand unification.

The outline of the paper is as follows. In section~\ref{Higgs} the
model will be introduced and the mass eigenvalues and eigenstates, as
well as the tree-level effective lagrangian, computed.  A (moderate)
$\mu$ problem is pointed out and a possible dynamical solution is
outlined. In section~\ref{EWSB} the conditions for EWSB are
presented. We will establish that EWSB will take place radiatively:
when there is a (partial) cancellation between the positive one-loop
gauge corrections and the (negative) tree-level masses EWSB is
triggered by the two-loop corrections induced by the top-stop
sector. The degree of fine-tuning is analyzed and proven to be much
less acute than in the MSSM. In section~\ref{DM} numerical solutions
are presented for a generic example and the typical supersymmetric
spectra are depicted. All bounds, from direct searches at LEP on Higgs
masses and from indirect electroweak precision observables, can be
satisfied for compactification scales $M_c\simgt 6$ TeV. We have also
studied the constraints from the requirement that the LSP annihilates
at a rate consistent with recent WMAP data which leads to very heavy
($\simgt 20$ TeV) gauginos and almost Dirac light Higgsinos, still
consistent with all experimental data. Finally in
section~\ref{conclusions} we present our conclusions.

\section{\sc The Higgs sector and tree level soft terms}
\label{Higgs}

The Higgs field is a 5D hypermultiplet which is a doublet under
$SU(2)_W\otimes SU(2)_H$ where $SU(2)_H$ is a global symmetry
introduced to account for two Higgs hypermultiplets. Although we will
break the latter symmetry by both bulk and brane mass terms, it is
useful to establish a covariant notation. We assume a flat extra
dimension with coordinate $y$ parametrizing the interval $0\leq y\leq
\pi$. We will work with 4D
superfields~\cite{Arkani-Hamed:2001tb,Marti:2001iw,arthur}. The
hypermultiplet can be written as two left handed chiral superfields
$\mathcal H, \mathcal H^c$ as
\be
\mathbb H^{a,i}=(\mathcal H, \bar {\mathcal H}^c)^{a,i}\,.
\ee
%H
Here $i$ and $a$ are the $SU(2)_W$ and $SU(2)_H$ indices
respectively. Note that the hypercharge assignment $Y=+\frac{1}{2}$
for $\mathbb H$ implies $Y=+\frac{1}{2}$ for $\mathcal H$ and
$Y=-\frac{1}{2}$ for $\mathcal H^c$.  The bulk Lagrangian for the
Higgs hypermultiplet reads:
\begin{multline}
\mathcal L^{\rm Higgs} = \int d^4\theta\ \frac{\mathcal T+\bar
{\mathcal T}}{2} \left\{\bar{\mathcal H} \, \exp(T_a V^a) \, \mathcal
H+ \mathcal H^{c}\, \exp(-T_aV^a)\, \bar{\mathcal H}^c\right\}\\ -\int
d^2\theta \left\{\mathcal H^c (\partial_y-\mathcal M \mathcal T)
\mathcal H+h.c.\right\}
\label{bulklag}
\end{multline}
where the mass matrix $\mathcal M$ is hermitian and in general
nondiagonal in $SU(2)_H$. We can parametrize it as
\be
\mathcal M=M'+M\, \vec p\cdot\vec \sigma\,.
\ee
where $M'$ and $M$ are arbitrary masses and $\vec p$ is a unit vector
in $su(2)_H$. The radion field $\mathcal T$ will be taken
nondynamical,
\be
\mathcal T=R+2\, \omega\,\theta^2 \,.
\ee
Its scalar component parametrizes the size of the extra dimension and
a non-zero $\omega$ implements the SS
breaking~\cite{Marti:2001iw,Kaplan:2001cg} (See also
Ref.~\cite{PaccettiCorreia:2004ri}).  The BC's for the fields
$\mathcal H$, $\mathcal H^c$ are determined by the variational
principle.  The boundary Lagrangian is taken to be
\be
\mathcal L_f^{\rm Higgs} = 
\int d^2 \theta\ \frac{1}{2}\left(
\mathcal H^c[1 +  \vec s_f\cdot\vec \sigma]\mathcal H+h.c.
\right)
\label{boundcov}
\ee
at the boundary at $y=y_f$ ($f=0,\,\pi$), and $\vec s_f$ is again a
unit vector in $su(2)_H$~\footnote{Eq.~(\ref{boundcov}) is a
sufficiently general boundary Lagrangian for our purposes. A more
detailed discussion of the most general boundary Lagrangian for the
system Eq.~(\ref{bulklag}) can be found in \cite{DGQ3}.}.  The form of
Eq.~(\ref{boundcov}) ensures consistent BC's by applying the
variational principle to the bulk+brane system defined in
Eqs.~(\ref{bulklag}) and (\ref{boundcov}), taking into account the
boundary terms which come from the variation of the term in
Eq.~(\ref{bulklag}) containing the derivative with respect to~$y$.

The superfield BC's following from this procedure are
\be
(1+\vec s_f\cdot\vec \sigma)\mathcal H=0,\qquad \mathcal H^c(1-\vec s_f\cdot\vec \sigma)=0
\label{BCs}
\ee
at the corresponding boundaries.  Note that the matrices acting on the
fields in Eq.~(\ref{BCs}) take the form of projectors such that some
linear combinations of $\mathcal H^a$ ($\mathcal H^c_a$) are set to
zero at the boundary, while the BC's of the ``orthogonal fields''
remain undetermined at this level. By means of a global $SU(2)_H$
rotation we can always achieve $\vec s_0=(0,0,-1)$ which will prove it
to be a convenient choice for us. With this convention, the BC's at
$y=0$ read:
\be
\mathcal H^2=0,\qquad \mathcal H^c_{1}=0.
\label{BCs0}
\ee
Note that this particular boundary condition ensures the absence of
potentially hazardous quadratically divergent FI terms as the two
chiral superfields $\mathcal H^1$ and $\mathcal H^c_2$, which do not
vanish at $y=0$ carry opposite hypercharge. These even
fields~\footnote{In analogy to the orbifold language, we will refer to
the fields which do not vansh at $y=0$ as ``even'' (at
$y=0$). Likewise the fields of Eq.~(\ref{BCs}) and Eq.~(\ref{BCs0})
will be called ``odd'' fields at $y=0$.} can be used to write Yukawa
superpotentials at this boundary for the up and down sectors
respectively:
\be W=\lambda_u \mathcal H^1(x,0) \mathcal Q(x)\mathcal U(x)
+\lambda_d \mathcal H^c_2(x,0)\mathcal Q(x)\mathcal D(x)\, ,
\label{yukawa}
\ee
where the 5D Yukawa couplings $\lambda_u$ and $\lambda_d$ have mass
dimension $-\frac{1}{2}$.

The BC's for the odd Higgs {\em scalars} $H^2$ and $H_1^c$ are given
by the $\theta=0$ component of Eq.~(\ref{BCs}), while those of the
even Higgs scalars $H^1$ and $H_2^c$ follow from computing the bulk
equations of motion (EOM) for the odd auxiliary fields and imposing
both the scalar and auxiliary component of the BC's
Eq.~(\ref{BCs}). At $y=0$ we find
\be
H^2=0,\qquad (\partial_y-M'+c_0M)H^1=0\,,
\label{BCscalar1}
\ee
\be
H^c_1=0,\qquad (\partial_y+M'+c_0M)H^c_2=0\,,
\label{BCscalar2}
\ee
where we define 
\be
c_f=\vec s_f\cdot\vec p,\qquad (f=0,\pi)\ .
\label{cf}
\ee 
At $y=\pi$, the equations take a similar form, replacing $H^{1,2}$ and
$H^c_{1,2}$ by the corresponding linear combinations.  Although our
boundary conditions avoid the generation of quadratically divergent FI
terms, there are linearly divergent contributions~\footnote{These
contributions can be seen to have their origin in the sign difference
between Eq.~(\ref{BCscalar1}) and Eq.~(\ref{BCscalar2}).} going as
$\sim\Lambda M'$.  To further reduce UV sensitivity in our model, we
will demand that $M'=0$ although these terms are much less dangerous
than the quadratically divergent ones.

We will assume that the $F$-term of the radion gets a VEV and triggers
SS-SUSY breaking~\cite{Marti:2001iw,Kaplan:2001cg}. The BC's determine
the mass spectrum through three SUSY preserving parameters: the angle
$\tilde \omega$, defined by
\be
\cos (2\pi \tilde \omega)=\vec s_0\cdot \vec s_\pi\ ,
\label{wt}
\ee
as well as the quantities $c_0$ and $c_\pi$ defined in
Eq.~(\ref{cf})~\footnote{Due to their geometric interpretation as
angles between vectors, these three parameters are not completely
independent. For fixed $\tilde \omega$, $c_0$ and $c_\pi$ must lie
within an elliptic disc~\cite{Diego:2005mu}. For $\tilde \omega=0\
(1/2)$ this disc degenerates to the line $c_0=c_\pi\ (-c_\pi)$.}.
Furthermore there is only one SUSY breaking parameter, the SS twist
$\omega\equiv |F_T/2|$.

The mass eigenvalues for the Higgs scalars are determined by the
zeroes of the equation~\cite{Diego:2005mu}
\be
\left(\cos(\Omega \pi R)-\frac{c_0 M}{\Omega}\sin(\Omega \pi R)\right)
\left(\cos(\Omega \pi R)+\frac{c_\pi M}{\Omega}\sin(\Omega \pi R)\right)
=\cos^2(\omega\pm\tilde\omega)\pi.
\label{masses}
\ee
where $\Omega^2=m^2-M^2$. For fermions (Higgsinos) we simply have to
set $\omega=0$ in Eq.~(\ref{masses}).  A detailed discussion of the
properties of Eq.~(\ref{masses}) can be found in
Refs.~\cite{Diego:2005mu,DGQ3}.  Here we will only consider an
interesting limit, that of quasilocalized fields. By assuming that $M
c_0>0$, for
\be
\epsilon\equiv \exp(-\pi c_0 M R)\ll 1
\ee
there are two 4D modes $H_\pm(x)$ whose wavefunctions localize towards
the boundary at $y=0$. They have masses
\be 
m_{\pm}^2/M^2 = (1-c_0^2) + 4
c_0^2\left(1-\frac{2\cos^2(\omega\pm\tilde\omega)\pi}{1+c_\pi/c_0}\right)
\epsilon^2+\mathcal O (\epsilon^4)\,.
\label{massesbos}
\ee
There might also be two modes localizing at $y=\pi$ which can be made
heavy~\cite{DGQ3}.  From now on we will only keep the two lightest
modes, which will make up the MSSM Higgs sector. The corresponding
Higgsinos have Dirac masses given by Eq.~(\ref{massesbos}) with
$\omega=0$. For the hyperscalars, the 5D fields can be approximated
as~\footnote{More precisely the corrections to Eq.~(\ref{wavefunc1})
and Eq.~(\ref{wavefunc2}) are $\mathcal O(\epsilon^{2-y/\pi})$, so for
$y=0$ the suppression is actually $\mathcal O(\epsilon^2)$ while at
$y=\pi$ it is $\mathcal O(\epsilon)$.}
\be
H^1(x,y)=\sqrt{ c_0 M} \exp(-c_0MRy)
\left[H_+(x)+H_-(x)\right]
+\mathcal O(\epsilon)
\label{wavefunc1}
\ee
\be
H_2^c(x,y)=\sqrt{ c_0 M} \exp(-c_0MRy)
\left[\bar H_-(x) - \bar H_+(x)\right]
+\mathcal O(\epsilon)
\label{wavefunc2}
\ee
while the dependence of $H^2(x,y)$ and $H_1^c(x,y)$ on $H_\pm(x)$ is
only of $\mathcal O(\epsilon)$.  From Eq.~(\ref{yukawa}) we can see
that the states
\be
H_u(x)\equiv\frac{1}{\sqrt2}\left(H_+ + H_-\right)
\label{up}
\ee
\be
H_d(x)\equiv\frac{1}{\sqrt2}\left(\bar H_- - \bar H_+\right)
\label{down}
\ee
can be identified with the MSSM up and down type Higgses respectively.
Note that the dimensionless 4D Yukawa couplings are given by
$y_{u,d}=\sqrt{2 c_0 M}\lambda_{u,d}$.

We can now easily calculate the tree-level (soft) masses in the
lagrangian
\be 
\mathcal L_{\rm mass}=-(\mu^2+m^2_{H_{u}})\,|H_u|^2
-(\mu^2+m^2_{H_{d}})\,
|H_d|^2+m^2_3\left(H_u\cdot H_d+h.c.\right)
\label{masas}
\ee
where $\mu$ is the Dirac mass of Higgsinos. The masses in
Eq.~(\ref{masas}) take the following general form
\be
\mu^2 = \left(s_0^2 +  a_1\,\epsilon^2\right)\, M^2\,,
\qquad s_0^2\equiv1-c_0^2\,,
\label{mutree}
\ee
\be
m_{H_u}^2=m_{H_d}^2=a_2\,\epsilon^2\,M^2\, ,
\label{mudtree}
\ee
\be
m_{3}^2=a_3\,M^2\,
\epsilon^2\,.
\label{btree}
\ee
where the coefficients $a_i$ are $\mathcal O(1)$ numbers which depend
on the BC parameters and all higher order corrections are $\mathcal
O(\epsilon^4)$.  Typically we would like all these masses to be of
$\mathcal O(m_Z)$ for EWSB to occur without too much
fine-tuning~\footnote{For a typical compactification scale
$M_c\equiv1/R\sim 5-10$ TeV, Eqs.~(\ref{mudtree}) and (\ref{btree})
would require $M\simgt M_c$, giving $\epsilon\sim 10^{-2}$. The fact
that $M\sim M_c$ is a generic prediction in this class of models and
might indicate that $M$ plays some role in the stabilization of $R$,
as for instance in~\cite{Luty:2002hj,vonGersdorff:2003rq}.}. We are
thus forced to choose $s_0=\mathcal O (\epsilon)$; for the geometry of
the BC's this means that the angle between the vectors $\vec s_0$ and
$\vec p$ is very small and as a consequence $c_0\simeq 1$ and
$c_\pi\simeq \cos(2\pi\tilde \omega)$.

The smallness of $s_0$ ($|s_0|\sim 1\%$ of the theoretically allowed
region $|s_0|\leq 1$) gives rise to a $\mu$-problem in our model at
the percent level. The fact that the $\mu$-term and the soft terms
arise at different orders in the $\epsilon$ expansion can be traced
back to the following fact. Notice that both boundary and bulk mass
matrices preserve $U(1)_H$ subgroups of the global $SU(2)_H$,
generated by $\vec s_f\cdot \vec\sigma$ and $\vec p\cdot\vec\sigma$
respectively.  For $\vec s_0=\pm \vec p$ (corresponding to $s_0=0$)
the surviving $U(1)$ at $y=0$ and the $U(1)$ in the bulk coincide,
this symmetry being broken only by the mismatched $U(1)$ at
$y=\pi$. The zero modes feel this breaking through their
wavefunctions, which are however suppressed at $y=\pi$ as $\sim
\epsilon$. Hence we expect $\mu\sim \epsilon^2$ when $s_0=0$ as it can
be checked from the $\epsilon$ expansion of fermionic mass
eigenvalues. When $s_0\neq0$, the breaking of the $U(1)$ at $y=0$ is
really felt to $\mathcal O(1)$ as it occurs even for infinitesimally
small $y>0$ and hence the $\mu$-term is unsuppressed.  On the other
hand supersymmetry is broken \`a la Scherk-Schwarz, which can be
interpreted as a mismatch of the surviving boundary $U(1)_R$ subgroups
of the $N=2$ $SU(2)_R$ automorphism group in the bulk. Again, the
zero-mode wave-functions feel this only to $\mathcal O(\epsilon)$, and
the corresponding soft terms are suppressed as $m^2\sim \epsilon^2$.

One could think of a dynamical solution to the $\mu$ problem in the
following way. Assume the relation $s_0=0$ to be exact at the 5D
cutoff scale $\Lambda$. If the resulting $U(1)$ symmetry at $y=0$ is
only broken by the VEV of a localized SM singlet field $S$ coupling
as~\footnote{The factor $\Lambda^{-1}$ has been introduced to render
the coupling dimensionless, and we assume it to be of $\mathcal
O(1)$.}
\be
\Delta W= \Lambda^{-1} \mathcal S(x) \mathcal H^1(x,0) \mathcal H^c_2(x,0)\,,
\ee
the $\mu$-term will be directly proportional to $\delta =\langle
S\rangle/ \Lambda$. The quantity $\delta$ will be small if the $U(1)$
breaks at a lower scale, i.e.~$\langle S\rangle \ll\Lambda$. In fact,
one can see that the backreaction of the new dynamically generated
$\mu$-term on the boundary conditions gives $s_0\simeq\delta$. Note
that this mechanism is the 5D version of the 4D NMSSM where a singlet
is coupled to the Higgs superfields as the term $\mathcal S\mathcal
H_u\mathcal H_d$ in the superpotential. While assuming $\mathcal O(1)$
Yukawa couplings, in 4D the VEV $\delta$ is constrained by EWSB to be
$\delta\sim m_Z/\Lambda$, in our 5D theory the much milder constraint
$\delta\sim m_Z/M_c$ holds. For the purpose of this paper we will
simply assume $s_0$ to be a small quantity.

In the approximation of small $s_0$ the soft terms become
\be
m_{H_u}^2=m_{H_d}^2= 4  M^2 \sin^2(\pi\omega)
(1-\tan^2(\pi\tilde \omega))\ \epsilon^2
\label{tree}
\ee
\be
m_{3}^2=4  M^2
\sin(2\pi\omega)\tan(\pi\tilde \omega) \
\epsilon^2
\ee
while the $\mu$-term is given by
\be
\mu^2 = s_0^2 M^2+\mathcal O(s_0^2\epsilon^2)\,.
\label{treeb}
\ee

Finally, we would like to comment on the quartic $D$-term
potential. As it has been shown in Ref.~\cite{Mirabelli:1997aj}, for a
gauged localized chiral multiplet the tree-level quartic potential
becomes proportional to $\delta(0)$. In other words this quartic
potential appears to violate SUSY by a huge amount. It was further
shown in~\cite{Mirabelli:1997aj} that due to a trilinear interaction
of the scalar with the adjoint chiral multiplet $\Sigma$, the $N=2$
partner of the bulk vector multiplet, these singularities cancel in
all physical processes. In our setup the Higgs is quasi-localized and
all delta functions are regularized with width $M^{-1/2}$, causing the
$D$-term to diverge as $\sim M$.  As it will be shown in~\cite{DGQ3},
one can calculate the effect of $\Sigma$ to the quartic potential
exactly by solving its 5D EOM.  This can be interpreted
diagrammatically as integrating out the heavy KK modes of $\Sigma$,
generating an effective $H^4$ interaction from the trilinear
$H^2\Sigma$ coupling.  Adding the latter to the explicit D-term we get
for the neutral components of the Higgs doublets
\be V_{\rm quartic}=\frac{g^2+g^{\prime\,2}}{8} (|H_u|^2-|H_d|^2)^2
+\mathcal O(\epsilon^2)\,.  \ee
where $g$ and $g^\prime$ are the 4D $SU(2)\otimes U(1)_Y$ gauge
couplings.  The leading contribution is thus precisely the MSSM one,
making explicit the cancellation of the divergent contributions in the
strictly localized limit which corresponds to $\epsilon\to 0$.

\section{\sc Electroweak symmetry breaking}
\label{EWSB}

In this section we will investigate in some detail the possibility of
EWSB. The conditions for EWSB and stability of the flat
$|H_u|=\pm|H_d|$ directions
\begin{align}
&(\mu^2+m_{H_u}^2)\,(\mu^2+m_{H_d}^2)<m_{3}^4\nonumber\\
&2\mu^2+m_{H_u}^2+m_{H_d}^2>2|m_{3}^2|
\end{align}
are incompatible with the tree-level induced SS supersymmetry breaking
where $m_{H_u}^2=m_{H_d}^2$. In this way EWSB should proceed
radiatively and we must incorporate radiative corrections to the Higgs
potential.  As matter is strictly localized and Higgses are
quasi-localized, SUSY breaking will predominantly be mediated by
one-loop gaugino loops that provide a (positive) contribution to the
squared masses of squarks, sleptons and Higgses.

In particular the squark masses will be dominated by the contribution
from the gluinos which is given by~\cite{PQ}
\be
\Delta m_{\tilde t,\tilde b}^2=
\frac{2\,g_3^2}{3\pi^4}\, M_c^2\,
f(\omega)
\label{mstop}
\ee
where the function $f(\omega)$ is defined by
\be 
f(\omega)\equiv \sum_{k=1}^\infty \frac{\sin(\pi k
\omega)^2}{k^3} \ ,
\label{fomega}
\ee
while electroweak gauginos provide a radiative correction to the
slepton and Higgs masses as
\be
\Delta^{(1)} m_{H_u}^2=\Delta^{(1)} m_{H_d}^2=
\frac{3g^2+g^{\prime\,2}}{8\pi^4}\,M_c^2
f(\omega)
\label{one}
\ee
Furthermore there is a sizable two-loop contribution to the Higgs soft
mass terms, as well as to the quartic coupling, coming from top-stop
loops with the one-loop generated squark masses given by
Eq.~(\ref{mstop}). This contribution can be estimated in the large
logarithm approximation by just plugging the one-loop squark masses in
the one-loop effective potential generated by the top-stop
sector~\cite{PQ}. The goodness of this approximation has been shown in
Ref.~\cite{Barbieri:2002uk} where a rigorous two-loop calculation of
the effective potential has been performed. For the sake of this
paper, where EWSB will not be marginal (as we will see later) it is
enough to consider the effective potential in the large logarithm
approximation, which yields the two-loop corrections to the Higgs
masses
\be
\Delta^{(2)} m_{H_u}^2=\frac{3 y_t^2}{8 \pi^2}\Delta m_{\tilde
t}^2 \, \log \frac{\Delta m_{\tilde t}^2}{\mathcal Q^2}
\,,
\label{twot}
\ee
\be
\Delta^{(2)} m_{H_d}^2=\frac{3 y_b^2}{8 \pi^2}\Delta m_{\tilde
b}^2 \, \log \frac{\Delta m_{\tilde t}^2}{\mathcal Q^2}
\,,
\label{twob}
\ee
where the renormalization scale should be fixed to the scale of SUSY
breaking, i.e.~the gaugino mass $\omega M_c$~\cite{PQ}. Notice that
the corrections from the bottom sector are also considered, which
would only be relevant for large values of $\tan \beta$.

A word has to be said about the bulk Higgs-Higgsino one-loop
contribution to the soft masses.  The reason we did neglect them with
respect to the one-loop gauge contribution (and even the leading
two-loop one) above is that they are strongly suppressed due to their
quasi-localization.  The leading $\mathcal O(\epsilon)$ corrections
come from the Higgs-Higgsino loop contribution to the stop mass. They
are proportional to the tree level soft Higgs mass $m_{H_u}^2\sim
M^2\epsilon^2 $ and hence suppressed as $\epsilon^2\log \epsilon$ with
respect to the gluon-gluino contribution of Eq.~(\ref{mstop}). We will
typically find values of $\epsilon\sim 10^{-2}$ and thus these
corrections are really subleading.  In principle we could easily
incorporate in our analysis the radiative corrections to $m_{3}^2$ as
calculated in~\cite{Delgado:2001si}.  However for most of the part of
parameter space we are interested in, this is only a tiny correction
to the tree level value, Eq.~(\ref{treeb}), and we will neglect it in
our analysis.

Finally, the leading two-loop corrections to the quartic self coupling
of $H_u$ and $H_d$ in the potential 
\be
\Delta V_{\rm quartic}= \Delta\gamma_u |H_u|^4+\Delta\gamma_u |H_u|^4 
\ee
are given by
\be
\Delta\gamma_{u}=\frac{3y_t^4}{16\pi^2}
\log\frac{\Delta m_{\tilde t}^2+m_t^2}{m_t^2}\,,
\ee
\be
\Delta\gamma_{d}=\frac{3y_b^4}{16\pi^2}
\log\frac{\Delta m_{\tilde b}^2+m_b^2}{m_b^2}\,.
\ee
where $m_t$ and $m_b$ are the the top and bottom quark masses
respectively.

Electroweak symmetry breaking can now occur in our model in a very
peculiar and interesting way. In fact the tree-level squared soft
masses $m_{H_u,H_d}^2$ given in Eq.~(\ref{tree}) are suppressed by the
factor $\epsilon^2$ and therefore, for values of $M\sim M_c$ they can
be comparable in size to the one-loop gauge corrections
$\Delta^{(1)}m_{H_u,H_d}^2$ given by Eq.~(\ref{one}).  Furthermore
the tree-level masses $m_{H_u,H_d}^2$ are negative for values of
$\tilde\omega>1/4$ and then there can be a (total or partial)
cancellation between the tree-level and one-loop contributions to the
Higgs masses. Under extreme conditions they can even cancel,
$m_{H_u,H_d}^2+\Delta^{(1)}m_{H_u,H_d}^2\simeq 0$, in which case the
negative two-loop corrections $\Delta^{(2)}m_{H_u}^2$ will easily
trigger EWSB. On the other hand in the limit of exact localization of
the Higgs fields $\epsilon\to 0$ the tree-level masses will vanish and
the one-loop gauge and two-loop top-stop corrections have to compete,
which will make the EWSB marginal, as pointed out in
Ref.~\cite{Barbieri:2002uk}. Similarly for $\tilde\omega\leq 1/4$ the
tree level masses $m_{H_u,H_d}^2$ are positive definite making it
again difficult, although not impossible as we argued in
section~\ref{introduction}, EWSB. These simple arguments prove that
there is a wide region in the space of parameters
$(\omega,\tilde\omega,\epsilon)$ where EWSB easily happens without any
fine-tuning of these parameters. Of course EWSB also depends on the
Higgsino mass $\mu$ and on the compactification scale $M_c$ (or
equivalently on the gluino mass as it happens in the MSSM) and we will
be concerned about the possible fine-tuning in those mass parameters.

It is easy to check that, due to the smallness of the SUSY breaking
scale which will be in the TeV region, as well as the extreme softness
of the SS mechanism, the usual fine-tuning problems of the MSSM can
almost entirely be avoided. To see this consider the $Z$ mass from the
minimization conditions of the potential in the limit $1\ll
\tan^2\beta\ll m_t^2/m_b^2$
\be \frac{m_Z^2}{2} =
-(\mu^2+m_{H_u}^2+\Delta^{(1)}m_{H_u}^2+\Delta^{(2)} m_{H_u}^2)\,.
\label{ft}
\ee
As it is intuitively clear, essentially no fine tuning is necessary if
we can make EWSB to work with all terms in Eq.~(\ref{ft}) roughly of 
electroweak size. Let us quantify a little further this statement by 
considering the sensitivy~\cite{Dimopoulos:1995mi} with respect to the 
fundamental parameters $M_i$
\be
\Delta_{M_i}=\left|\frac{M_i^2}{m_Z^2}\,
\frac{\partial m_Z^2}{\partial M_i^2}\right|
\ee
where $M_i=\mu,m_{H_u},M_c$~\footnote{We are defining our fundamental
parameters such that the sensitivity on them is really a measure of
fine-tuning in the sense of Ref.~\cite{fine}.}.  In terms of these
fundamental parameters Eq.~(\ref{ft}) can be rewritten as
\be
m_Z^2=-2\mu^2-2 m_{H_u}^2-\kappa M_c^2
\ee
where typically $\kappa\sim 10^{-3}$, and the corresponding
sensitivity parameters are given by
\begin{align}
\Delta_\mu&=\frac{2\mu^2}{m_Z^2}\nonumber\\
\Delta_{M_c}&=|\kappa|\,\frac{M_c^2}{m_Z^2}\nonumber\\
\Delta_{M_{H_u}}&=\left|1+\Delta_\mu+{\rm sign}(\kappa)\Delta_{M_c}\right|
\label{sensiparam}
\end{align}

\begin{figure}[htb]
\begin{center}
\includegraphics[width=8cm]{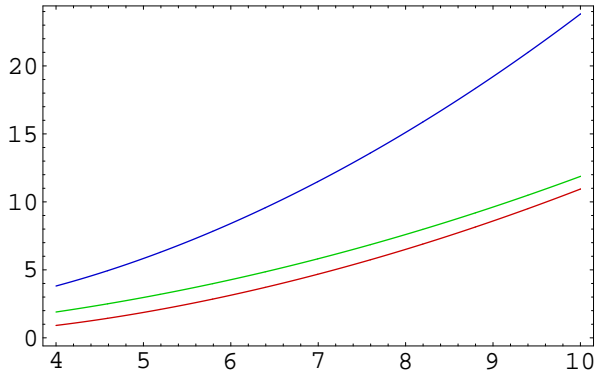}
\caption{\it The sensitivity parameters in Eq.~(\ref{sensiparam}) as
functions of $M_c$ in TeV for the case $\omega=0.45$,
$\tilde\omega=0.35$ and $M=1.65 M_c$.  From top to bottom the lines
are: $\Delta_{m_{H_u}}$ (blue line), $\Delta_{M_c}$ (green line) and
$\Delta_\mu$ (red line).}
\label{finetuning}
\end{center}
\end{figure}

In Fig.~\ref{finetuning} we plot the three sensitivity parameters in
(\ref{sensiparam}) for the model, that we will present in
section~\ref{DM}, corresponding to $\omega=0.45$, $\tilde\omega=0.35$
and $M=1.65 M_c$. This model gives a viable spectrum and it is
consistent with all electroweak precision observables for $M_c\simgt
6.5$ TeV.
As one sees from Fig.~\ref{finetuning} and Eq.~(\ref{sensiparam}) the
largest sensitivity appears to be with respect to the parameter
$m_{H_u}$. In fact for $M_c=6.6$ TeV the required amount of
fine-tuning is $\sim 10\%$ while for larger values of $M_c$ the
fine-tuning naturally increases quadratically. Thus for instance for
$M_c=10$ TeV the fine-tuning is $\sim 4\%$

We can now compare this situation with the one in the MSSM. The gluino
mass for a given value of $M_c$ is $M_3=\omega M_c$ so that in our
example, for $M_c\sim 10$ TeV we have $M_3\sim 5$ TeV.  In the MSSM
the $Z$ mass squared is proportional to $M_3^2$ for the same reason as
in our model, but with a much larger coefficient $\mathcal O(1)$ due
to large logarithms $\log m_Z/m_{\rm GUT}$.  A gluino of mass a few
TeV in the MSSM will require a ($\tan\beta$ dependent) fine-tuning as
large as 0.01\%.  A careful treatment of the fine tuning issues
related to the gluino mass can be found in
Ref.~\cite{Wright:1998mk}. This back-of-an-envelope calculation just
wanted to stress the rough differences between our mechanism of EWSB
and typical results in the MSSM.

\section{\sc Supersymmetric spectra and Dark Matter}
\label{DM}

We will now calculate the Higgs and superpartner spectra for some
specific values of the parameters. We would like to plot our
predictions as functions of $M_c$ with all other parameters
$(\omega,\tilde\omega,M)$ fixed. Because of the exponential dependence
of the tree level soft masses it will prove convenient to trade $M$ by
$\epsilon$ (which provides a fixed ratio of $M/M_c$) when varying over
$M_c$ in order to avoid excessively large or small masses.

The parameters $\omega$ and $\tilde \omega$ give $\mathcal O(1)$
coefficients in the soft parameters.  Their possible values can be
further restricted by demanding that the right-handed slepton mass
$m_{\tilde e_R}$ be above the mass of the lightest neutralino, as
there are strong constraints on charged stable
particles~\cite{Eidelman:2004wy} and we would like the lightest
neutralino to be the lightest supersymmetric particle (LSP) and a Dark
Matter candidate. For the nature of the latter notice that gaugino
masses are given by $\omega M_c$ while Higgsino masses are essentially
controlled by the $\mu$-parameter. We thus expect the neutralino to be
almost pure Higgsino with a mass basically given by $\mu$. On the
other hand the right handed slepton mass is radiatively generated and
proportional to $g' M_c$. The size of the $\mu$ term is determined by
the minimization conditions and will increase $\sim M_c$ for large
$M_c$ (as it has to compensate the negative radiative corrections to
$m_{H_u}^2$). However the tree level soft mass terms Eq.~(\ref{tree})
increase for smaller $\tilde \omega$ which in turn allows for a
smaller $\mu$. The requirement that the neutralino be lighter than the
charged sleptons thus favours the region $\omega >\tilde \omega$.

We then solve the minimization conditions for EWSB which will give us
two predictions, $\tan\beta$ and $\mu$ as functions of the only left
free parameter, $M_c$. Then all masses will become functions of
$M_c$. In particular in the Higgs sector all masses are obtained from
the effective potential where the one-loop corrections to the quartic
couplings are included. The mass of the SM-like Higgs is then computed
with radiative corrections to the quartic couplings considered at the
one-loop level. It is well known that including just the one-loop
effective potential overestimates somehow the Higgs masses and
improving the effective potential by an RGE resummation of leading
logarithms provides more realistic results. In this paper we will
nevertheless be content by evaluating masses in the one-loop
approximation. The squark and slepton masses are dominated by the
gaugino loop contribution and hence grow approximately linearly with
$M_c$.  We find~\cite{PQ}
\be 
(m_{\tilde
q_L},\,m_{\tilde u_R},\,m_{\tilde d_R},\,m_{\tilde \ell_L},\,m_{\tilde
e_R}) =(0.110,\,0.103,\,0.102,\,0.042,\,0.025)\sqrt{f(\omega)}M_c 
\label{relacion}
\ee
where the function $f(\omega)$ is given in
Eq.~(\ref{fomega})~\footnote{Numerically $f(\omega)\simlt 1$ for the
values of $\omega$ we will be interested in.}.

On the other hand the gauginos have a mass given by
\be
M_{1/2}=\omega M_c\ ,
\ee
and the Higgsinos, charginos and neutralinos, a mass approximately
equal to $\mu$, $m_{\tilde{\chi}^\pm}\simeq m_{\tilde{\chi}^0}\simeq
\mu$. They are quasi-degenerate in mass and its mass difference can be
given to a very good approximation (for $\mu<0$)
by~\cite{Giudice:1995qk}
\be 
\frac{\Delta
m_{\tilde{\chi}}}{m_W}\equiv\frac{m_{\tilde{\chi}^\pm}-m_{\tilde{\chi}^0}}{m_W}
\simeq (0.35+0.65 \sin 2\beta)\frac{m_W}{M_{1/2}} 
\ee
which means that typically e.g.~for $M_c\sim 10$ TeV, $\Delta
m_{\tilde\chi}\sim 1$ GeV. The phenomenology for Tevatron and $e^+e^-$
colliders of models where charginos and neutralinos are
quasi-degenerate in mass was worked out in Refs.~\cite{Gunion:1999jr}.
The most critical ingredients in the phenomenology of these models are
the lifetime and decay modes of $\tilde{\chi}^\pm$ which in turn
depend almost entirely on $\Delta m_{\tilde\chi}$. Conventional
detection of sparticles is difficult since the decay products
($\tilde{\chi}^\pm\to \tilde{\chi}^0\pi^\pm,
\tilde{\chi}^0\ell^\pm\nu_\ell,\dots$) are very soft and alternative
signals must be considered~\cite{Gunion:1999jr}.
                                                                     
\begin{figure}[htb]
\includegraphics[width=7.5cm]{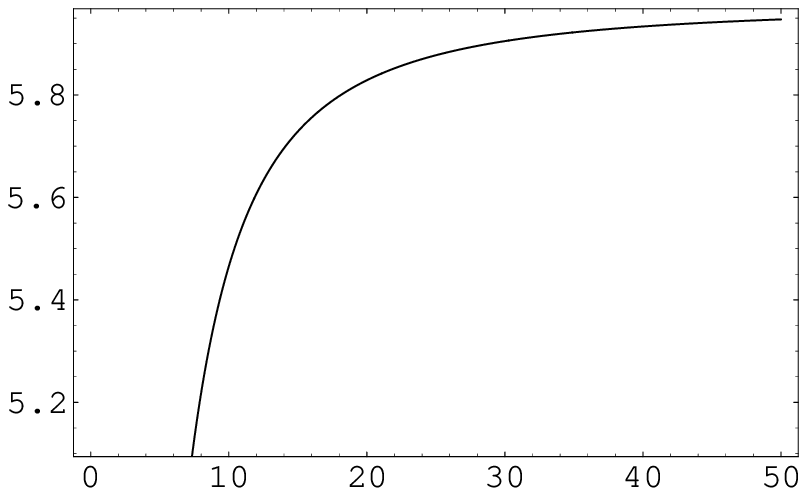}
\hspace{0.5cm}
\includegraphics[width=7.25cm]{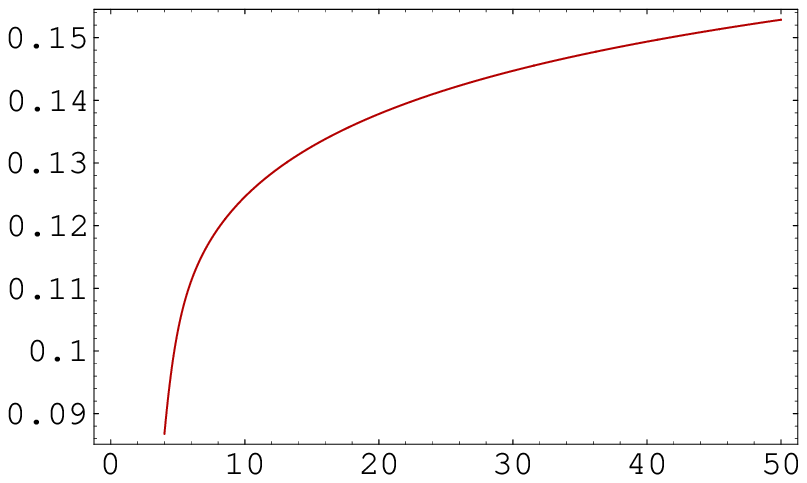}\\\vspace{0.25cm}
\includegraphics[width=7.25cm]{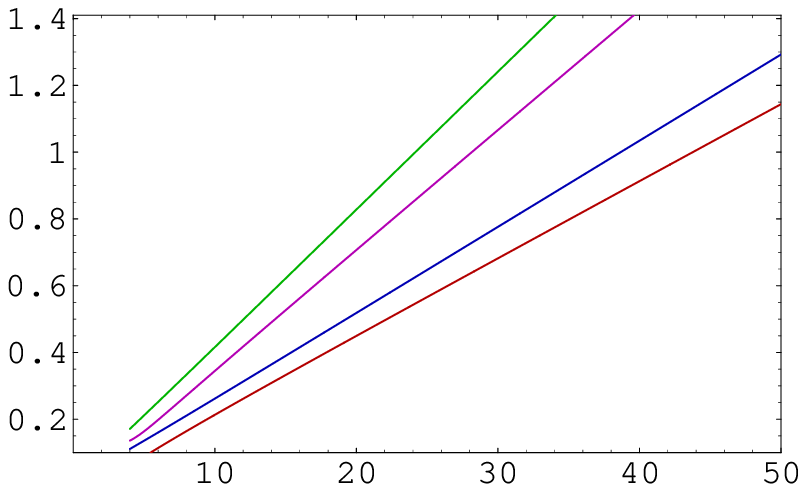}
\hspace{0.9cm}
\includegraphics[width=7.cm]{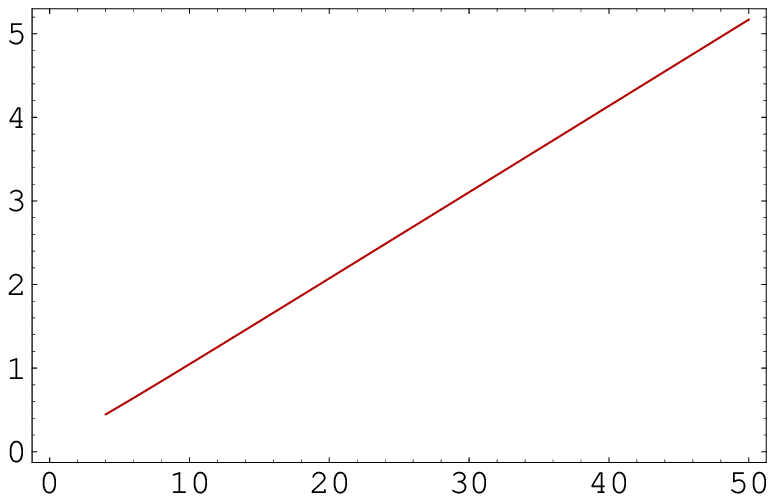}
\caption{\it Predictions for the case $\omega=0.45$,
$\tilde\omega=0.35$, $M=1.65M_c$ (as in Fig.~\ref{finetuning}) as a
function of the compactification scale. Upper left panel:
$\tan\beta$. Upper right panel: the SM-like Higgs mass $m_h$. Lower
left panel, from top to bottom the lines correspond to the masses of:
left-handed sleptons $m_{\tilde\ell_L}$ (green line), heavy neutral
Higgs (with a mass approximately equal to the pseudoscalar mass)
$m_H\simeq m_A$ (magenta line), right-handed sleptons $m_{\tilde e_R}$
and neutralinos $m_{\chi^0}\simeq \mu$ (red line). Lower right panel:
the squark masses $m_{\tilde q}$. All masses are in TeV.}
\label{espectro}
\end{figure}

We will now consider in detail a typical example that will be solved
numerically and we will plot all the predictions of the model as
functions of $M_c$. We choose $\omega=0.45$, $\tilde\omega=0.35$ and
$M=1.65 M_c$ as in the previous example of Fig.~\ref{finetuning} where
the fine-tuning in these models is exemplified. The results are shown
in Fig.~\ref{espectro}.
The SM-like Higgs mass easily satisfies the experimental bound
$m_{h^0}>114.5$ GeV for $M_c>6.5$ TeV. The LSP is the Higgsino-like
with mass $\sim\mu$.  Electroweak precision observables also put lower
bounds on $M_c$ (see e.g.~Ref.~\cite{Delgado:1999sv}). For the
particularly chosen model the $\chi^2(M_c)$ distribution has a minimum
around $M_c\simeq 10.5$ TeV and one deduces $M_c>4.9$ TeV at 95\% c.l.

\begin{figure}[htb]
\begin{center}
\includegraphics[width=8cm]{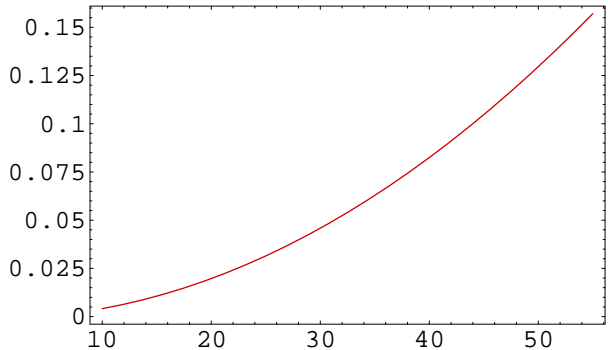}
\caption{\it $\Omega_{\tilde{\chi}^0} h^2$ as a function of $M_c$ (in TeV) for
the model presented in Fig.~\ref{espectro}.}
\label{Omega}
\end{center}
\end{figure}

Finally in the considered class of models where the neutralino is the
LSP and $R$-parity is conserved the lightest neutralino is the
candidate to Cold Dark Matter. In fact the prediction of
$\Omega_{\tilde{\chi}^0} h^2$ can be obtained using the DarkSUSY
package~\cite{Gondolo:2002tz} and can also be approximated by the
expression~\cite{Giudice:2004tc}
\be
\Omega_{\tilde{\chi}^0} h^2\simeq 0.09\left(\mu/TeV\right)^2
\ee
In the particular model of Fig.~\ref{espectro} the prediction of
$\Omega_{\tilde{\chi}^0} h^2$ is given in Fig.~\ref{Omega}

Recent WMAP results~\cite{WMAP} imply that
$0.114<\Omega_{\tilde{\chi}^0} h^2<0.134$. As one can see from
Fig.~\ref{Omega} this range in $\Omega_{\tilde{\chi}^0} h^2$ points
towards the range~\footnote{Of course, such large values of $M_c$
require a fine tuning $< 1\%$, see section.~\ref{EWSB}.} $49\, {\rm
TeV}<M_c<53\,{\rm TeV}$. Then for a value of $M_c\sim 50$ TeV the
density of Dark Matter agrees with the recent results obtained from
WMAP. Notice that for such large values of $M_c$ the neutralinos are
almost Dirac particles. However the non-Diracity is spoiled by
$\mathcal O(m_W/M_{1/2}) m_W\sim 300$ MeV which is enough to avoid the
strong limits on Dirac fermions that put a lower bound on the
non-Diracity around 100 KeV~\cite{diracity}. On the other hand the
WMAP range for $M_c$ implies, in the gravitational sector, gravitino
masses $m_{3/2}\simgt 10$ TeV (depending on the value of the SS
parameter $\omega$) are such that gravitinos decay early enough to
avoid cosmological troubles and thus solving the longstanding
cosmological gravitino problem~\cite{Weinberg:1982zq}.

\section{\sc Conclusions}
\label{conclusions}

In this paper we have analyzed electroweak symmetry breaking in a
five-dimensional model where supersymmetry is broken by Scherk-Schwarz
boundary conditions and quark and lepton superfields are localized at
one of the boundaries. The gauge sector propagates in the bulk and
thus gauginos receive tree-level masses from the Scherk-Schwarz
mechanism (they are the heaviest supersymmetric particles) while
squarks and sleptons acquire one-loop supersymmetry breaking masses
from the bulk (they can be heavy but lighter than gauginos). The Higgs
squared masses receive positive one-loop contributions from the gauge
sector and negative two-loop contributions from the top-stop sector,
the latter applying both to Higgses which belong to localized
multiplets on the boundary and to zero modes of Higgses which belong
to hypermultiplets propagating in the bulk. Under these circumstances
negative two-loop corrections have to compete with positive one-loop
effects and therefore electroweak breaking is marginal if not
impossible.

If Higgses propagate in hypermultiplets in the bulk, but if they are
quasi-localized by a supersymmetric mass, they feel the Scherk-Schwarz
supersymmetry breaking as they are bulk fields but, on the other hand,
their mass is not only controlled by the compactification radius $1/R$
but also by the localizing mass $M$. In fact in the "localization
limit", where $\epsilon=\exp(-\pi MR)\ll 1$, the squared masses can be
comparables in size with the radiatively generated ones. Furthermore
those squared tree-level masses can be physical (positive), tachyonic
(negative) or even zero. The situation concerning electroweak symmetry
breaking is thus very peculiar and interesting:

\begin{itemize}
\item
If the tree-level masses are physical (or zero) electroweak breaking
should be triggered at two-loop, which makes it marginal as we already
pointed out. Notice that the case of localized Higgses corresponds to
the limit $\epsilon\to 0$ where the tree-level soft masses are zero
while a finite supersymmetric mass ($\mu$ term) may remain.
\item
If tree-level masses are tachyonic the conditions for electroweak
symmetry breaking with stable $D$-flat directions are incompatible to
each other, since the generated tree-level masses of the two Higgses
are equal. However the introduction of radiative corrections, that
discriminate between the $H_u$ and $H_d$ masses through the
corresponding Yukawa couplings can trigger electroweak symmetry
breaking.
\end{itemize}
Therefore electroweak breaking is neither purely triggered by
tree-level masses nor by radiative corrections but both effects are
needed: we could dub it as {\it tree-level assisted electroweak
radiative breaking}.

The main features of these models can be summarized as follows:

$\bullet$ No quadratically divergent Fayet-Iliopoulos terms do appear
so the Higgs mass is one-loop finite.

$\bullet$ Gauginos are the heaviest supersymmetric particles (they are
in the TeV or multi-TeV region). Supersymmetry breaking is mediated by
gauginos to the observable sector and flavor-changing neutral currents
are naturally suppressed. Models are of "no-scale" type and then no
anomaly mediated supersymmetry breaking occurs.

$\bullet$ Squarks and sleptons acquire radiative masses from loops of
gluinos and electroweak gauginos, respectively. Their masses are then
suppressed with respect to those of gauginos by loop
factors. Furthermore, there is a striking prediction for the ratios of
sfermion masses Eq.~(\ref{relacion}). Note that similar relations are
known from gauge mediation models \cite{gaugemediation} (see
Ref.~\cite{Giudice:1998bp} for a review). There however scalar masses
are generated at the two loop level and hence different ratios
apply~\footnote{Depending on the size of the messenger scale and other
details of the model, these relations can receive important
corrections from RG running. In our case we expect RG effects to be
small, as the high scale ($M_c$) is only about two orders of
magnitude above the low scale ($m_Z$).}.

$\bullet$ Due to the smallness of the supersymmetry breaking scale (in
the TeV region) and the extreme softness of the Scherk-Schwarz
mechanism the fine-tuning problems of the MSSM can almost entirely be
avoided. For instance in our model gluinos around 3 TeV mass require a
modest 10\% fine-tuning.

$\bullet$ Higgsinos are the lightest supersymmetric particles (with a
mass in the sub-TeV region). Charged and neutral Higgsinos are almost
degenerate with mass splittings $\simlt~1$~GeV.

$\bullet$ The lightest supersymmetric particle is a neutralino which
is a good candidate to Dark Matter if its mass is around the TeV. This
would require multi-TeV gluinos (and gravitinos) that would in turn
require less than 1\% fine-tuning. The gravitinos will decay early
enough to avoid any cosmological problems.

The phenomenology of these models is also very peculiar. Since
gauginos are superheavy they might not be detected at LHC or
ILC. However since squarks are much lighter than gluinos the latter
could easily decay into squarks and quarks: gluinos are then
short-lived and thus they do not generate any cosmological
problems. The effective theory below the TeV scale is thus a two Higgs
doublet model with degenerate corresponding neutral and charged
Higgsinos and (left and right-handed) sleptons. Even if Kaluza-Klein
excitations might be too heavy for discovery at LHC there is a smoking
gun in this model: the mass ratio between different supersymmetric
particles is fixed by the relation of Eq.~(\ref{relacion}). The
phenomenology of these models is very sensitive to the mass difference
between charginos and neutralinos, which can be an indirect measure of
the Kaluza-Klein masses.

\vspace*{7mm}
\subsection*{\sc Acknowledgments}

\noindent This work was partly supported by CICYT, Spain, under
contracts FPA 2004-02012 and FPA 2005-02211. GG was supported by the
Leon Madansky Fellowship and by DOE grant DE-FG02-3ER41271.


\begin{thebibliography}{99}
%
\bibitem{Antoniadis:1990ew} I.~Antoniadis,
%``A Possible New Dimension At A Few Tev,''
Phys.\ Lett.\ B {\bf 246} (1990) 377.
%%CITATION = PHLTA,B246,377;%%
%

\bibitem{Quiros:2003gg}
M.~Quiros,
%``New ideas in symmetry breaking,''
arXiv:hep-ph/0302189.
%%CITATION = HEP-PH 0302189;%%
%

\bibitem{Randall:1998uk}
L.~Randall and R.~Sundrum,
%``Out of this world supersymmetry breaking,''
Nucl.\ Phys.\ B {\bf 557} (1999) 79
[arXiv:hep-th/9810155].
%%CITATION = HEP-TH 9810155;%%
%

\bibitem{Uranga:2005wn}
A.~M.~Uranga,
%``Intersecting brane worlds,''
Class.\ Quant.\ Grav.\  {\bf 22} (2005) S41;
%%CITATION = CQGRD,22,S41;%%
R.~Blumenhagen, M.~Cvetic, P.~Langacker and G.~Shiu,
%``Toward realistic intersecting D-brane models,''
arXiv:hep-th/0502005;
%%CITATION = HEP-TH 0502005;%%
I.~Antoniadis, K.~Benakli, A.~Delgado, M.~Quiros and M.~Tuckmantel,
%``Split extended supersymmetry from intersecting branes,''
arXiv:hep-th/0601003.
%%CITATION = HEP-TH 0601003;%%

%
\bibitem{Scherk:1978ta}
J.~Scherk and J.~H.~Schwarz,
%``Spontaneous Breaking Of Supersymmetry Through Dimensional Reduction,''
Phys.\ Lett.\ B {\bf 82} (1979) 60;
%%CITATION = PHLTA,B82,60;%%
%
J.~Scherk and J.~H.~Schwarz,
%``How To Get Masses From Extra Dimensions,''
Nucl.\ Phys.\ B {\bf 153} (1979) 61.
%%CITATION = NUPHA,B153,61;%%
%

\bibitem{Delgado:2001ex}
A.~Delgado, G.~von Gersdorff, P.~John and M.~Quiros,
%``One-loop Higgs mass finiteness in supersymmetric Kaluza-Klein theories,''
Phys.\ Lett.\ B {\bf 517} (2001) 445
[arXiv:hep-ph/0104112];
%%CITATION = HEP-PH 0104112;%%
%
A.~Delgado, G.~von Gersdorff and M.~Quiros,
%``Two-loop Higgs mass in supersymmetric Kaluza-Klein theories,''
Nucl.\ Phys.\ B {\bf 613} (2001) 49
[arXiv:hep-ph/0107233];
%%CITATION = HEP-PH 0107233;%%
%
R.~Contino and L.~Pilo,
%``A note on regularization methods in Kaluza-Klein theories,''
Phys.\ Lett.\ B {\bf 523} (2001) 347 
[arXiv:hep-ph/0104130].
%%CITATION = HEP-PH 0104130;%% 
N.~Weiner, %``Ineffective supersymmetry: Electroweak symmetry breaking from extra %dimensions,'' 
arXiv:hep-ph/0106021;
%%CITATION= HEP-PH 0106021;%% 
V.~Di Clemente and Y.~A.~Kubyshin, 
%``Effective potential and KK-renormalization scheme in a 5D supersymmetric
%theory,'' 
Nucl.\ Phys.\ B {\bf 636} (2002) 115
[arXiv:hep-th/0108117];
%%CITATION = HEP-TH 0108117;%% H.~D.~Kim,
H.~D.~Kim,
%``Softness of Scherk-Schwarz supersymmetry breaking,''
Phys.\ Rev.\ D {\bf 65} (2002) 105021
[arXiv:hep-th/0109101].
%%CITATION = HEP-TH 0109101;%%
%

\bibitem{PQ}
A.~Pomarol and M.~Quiros,
%``The standard model from extra dimensions,''
Phys.\ Lett.\ B {\bf 438} (1998) 255
[arXiv:hep-ph/9806263];
%%CITATION = HEP-PH 9806263;%%
I.~Antoniadis, S.~Dimopoulos, A.~Pomarol and M.~Quiros,
%``Soft masses in theories with supersymmetry breaking by
%TeV-compactification,''
Nucl.\ Phys.\ B {\bf 544} (1999) 503
[arXiv:hep-ph/9810410];
%%CITATION = HEP-PH 9810410;%%
A.~Delgado, A.~Pomarol and M.~Quiros,
%``Supersymmetry and electroweak breaking from extra dimensions at the
%TeV-scale,''
Phys.\ Rev.\ D {\bf 60} (1999) 095008
[arXiv:hep-ph/9812489].
%%CITATION = HEP-PH 9812489;%%
%

\bibitem{Barbieri:2002uk}
R.~Barbieri, G.~Marandella and M.~Papucci,
%``Breaking the electroweak symmetry and supersymmetry by a compact extra
%dimension,''
Phys.\ Rev.\ D {\bf 66} (2002) 095003
[arXiv:hep-ph/0205280];
%%CITATION = HEP-PH 0205280;%%
%
%\bibitem{Barbieri:2002sw}
R.~Barbieri, L.~J.~Hall, G.~Marandella, Y.~Nomura, T.~Okui, S.~J.~Oliver and M.~Papucci,
%``Radiative electroweak symmetry breaking from a quasi-localized top
%quark,''
Nucl.\ Phys.\ B {\bf 663} (2003) 141
[arXiv:hep-ph/0208153];
%%CITATION = HEP-PH 0208153;%%
%
R.~Barbieri, G.~Marandella and M.~Papucci,
%``The Higgs mass as a function of the compactification scale,''
Nucl.\ Phys.\ B {\bf 668} (2003) 273
[arXiv:hep-ph/0305044].
%%CITATION = HEP-PH 0305044;%%
%

\bibitem{Barbieri:2000vh}
R.~Barbieri, L.~J.~Hall and Y.~Nomura,
%``A constrained standard model from a compact extra dimension,''
Phys.\ Rev.\ D {\bf 63} (2001) 105007
[arXiv:hep-ph/0011311].
%%CITATION = HEP-PH 0011311;%%
%

\bibitem{Delgado:2001si}
A.~Delgado and M.~Quiros,
%``Supersymmetry and finite radiative electroweak breaking from an extra
%dimension,''
Nucl.\ Phys.\ B {\bf 607}, 99 (2001)
[arXiv:hep-ph/0103058].
%%CITATION = HEP-PH 0103058;%%
%
\bibitem{Marti}
D.~Marti and A.~Pomarol,
%``Fayet-Iliopoulos terms in 5d theories and their phenomenological
%implications,''
Phys.\ Rev.\ D {\bf 66} (2002) 125005
[arXiv:hep-ph/0205034].
%%CITATION = HEP-PH 0205034;%%

%
\bibitem{Delgado:1999sv}
A.~Delgado, A.~Pomarol and M.~Quiros,
%``Electroweak and flavor physics in extensions of the standard model with
%large extra dimensions,''
JHEP {\bf 0001} (2000) 030
[arXiv:hep-ph/9911252].
%%CITATION = HEP-PH 9911252;%%
%

%
\bibitem{Ghilencea:2001bw}
D.~M.~Ghilencea, S.~Groot Nibbelink and H.~P.~Nilles,
%``Gauge corrections and FI-term in 5D KK theories,''
Nucl.\ Phys.\ B {\bf 619} (2001) 385
[arXiv:hep-th/0108184].
%%CITATION = HEP-TH 0108184;%%%

%
\bibitem{Georgi:2000wb}
H.~Georgi, A.~K.~Grant and G.~Hailu,
%``Chiral fermions, orbifolds, scalars and fat branes,''
Phys.\ Rev.\ D {\bf 63} (2001) 064027
[arXiv:hep-ph/0007350];
%%CITATION = HEP-PH 0007350;%%
%
N.~Arkani-Hamed, A.~G.~Cohen and H.~Georgi,
%``Anomalies on orbifolds,''
Phys.\ Lett.\ B {\bf 516} (2001) 395
[arXiv:hep-th/0103135];
%%CITATION = HEP-TH 0103135;%%
%
S.~Groot Nibbelink, H.~P.~Nilles and M.~Olechowski,
%``Spontaneous localization of bulk matter fields,''
Phys.\ Lett.\ B {\bf 536} (2002) 270
[arXiv:hep-th/0203055].
%%CITATION = HEP-TH 0203055;%%

%
\bibitem{Diego:2005mu}
D.~Diego, G.~von Gersdorff and M.~Quiros,
%``Supersymmetry and electroweak breaking in the interval,''
JHEP {\bf 0511}, 008 (2005)
[arXiv:hep-ph/0505244].
%%CITATION = HEP-PH 0505244;%%

%
\bibitem{Hebecker:2002re}
A.~Hebecker and J.~March-Russell,
%``The flavour hierarchy and see-saw neutrinos from bulk masses in 5d
%orbifold GUTs,''
Phys.\ Lett.\ B {\bf 541} (2002) 338
[arXiv:hep-ph/0205143].
%%CITATION = HEP-PH 0205143;%%
%

%
\bibitem{Abe:2004tq}
H.~Abe, K.~Choi, K.~S.~Jeong and K.~i.~Okumura,
%``Scherk-Schwarz supersymmetry breaking for quasi-localized matter fields
%and supersymmetry flavor violation,''
JHEP {\bf 0409} (2004) 015
[arXiv:hep-ph/0407005].
%%CITATION = HEP-PH 0407005;%%

%
\bibitem{Ellis:1983sf}
J.~R.~Ellis, A.~B.~Lahanas, D.~V.~Nanopoulos and K.~Tamvakis,
%``No - Scale Supersymmetric Standard Model,''
Phys.\ Lett.\ B {\bf 134} (1984) 429;
%%CITATION = PHLTA,B134,429;%%
% 
J.~R.~Ellis, C.~Kounnas and D.~V.~Nanopoulos,
%``No Scale Supersymmetric Guts,''
Nucl.\ Phys.\ B {\bf 247} (1984) 373.
%%CITATION = NUPHA,B247,373;%%

%
%\cite{Luty:2002hj}
\bibitem{Luty:2002hj}
M.~A.~Luty and N.~Okada,
%``Almost no-scale supergravity,''
JHEP {\bf 0304} (2003) 050
[arXiv:hep-th/0209178];
%%CITATION = HEP-TH 0209178;%%

%
\bibitem{dudas}
E.~Dudas and M.~Quiros,
%``Five-dimensional massive vector fields and radion stabilization,''
Nucl.\ Phys.\ B {\bf 721} (2005) 309
[arXiv:hep-th/0503157].
%%CITATION = HEP-TH 0503157;%%

%
\bibitem{Mirabelli:1997aj}
E.~A.~Mirabelli and M.~E.~Peskin,
%``Transmission of supersymmetry breaking from a 4-dimensional boundary,''
Phys.\ Rev.\ D {\bf 58} (1998) 065002
[arXiv:hep-th/9712214];
%%CITATION = HEP-TH 9712214;%%

%
\bibitem{gauginomediation}
%\bibitem{Kaplan:1999ac}
D.~E.~Kaplan, G.~D.~Kribs and M.~Schmaltz,
%``Supersymmetry breaking through transparent extra dimensions,''
Phys.\ Rev.\ D {\bf 62}, 035010 (2000)
[arXiv:hep-ph/9911293];
%%CITATION = HEP-PH 9911293;%%
Z.~Chacko, M.~A.~Luty, A.~E.~Nelson and E.~Ponton,
%``Gaugino mediated supersymmetry breaking,''
JHEP {\bf 0001} (2000) 003
[arXiv:hep-ph/9911323].
%%CITATION = HEP-PH 9911323;%%
%

%
\bibitem{Dienes:1998vh}
K.~R.~Dienes, E.~Dudas and T.~Gherghetta,
%``Extra spacetime dimensions and unification,''
Phys.\ Lett.\ B {\bf 436} (1998) 55
[arXiv:hep-ph/9803466];
%%CITATION = HEP-PH 9803466;%%
%\bibitem{Dienes:1998vg}
%  K.~R.~Dienes, E.~Dudas and T.~Gherghetta,
%``Grand unification at intermediate mass scales through extra dimensions,''
Nucl.\ Phys.\ B {\bf 537} (1999) 47
[arXiv:hep-ph/9806292].
%%CITATION = HEP-PH 9806292;%%

%
\bibitem{Seiberg:1996bd}
N.~Seiberg,
%``Five dimensional SUSY field theories, non-trivial fixed points and  string
%dynamics,''
Phys.\ Lett.\ B {\bf 388} (1996) 753
[arXiv:hep-th/9608111];
%%CITATION = HEP-TH 9608111;%%
%\bibitem{Intriligator:1997pq}
K.~A.~Intriligator, D.~R.~Morrison and N.~Seiberg,
%``Five-dimensional supersymmetric gauge theories and degenerations of
%Calabi-Yau spaces,''
Nucl.\ Phys.\ B {\bf 497} (1997) 56
[arXiv:hep-th/9702198].
%%CITATION = HEP-TH 9702198;%%


%
\bibitem{Hebecker:2002vm}
A.~Hebecker and A.~Westphal,
%``Power-like threshold corrections to gauge unification in extra
%dimensions,''
Annals Phys.\  {\bf 305}, 119 (2003)
[arXiv:hep-ph/0212175];
%%CITATION = HEP-PH 0212175;%%
%\bibitem{Hebecker:2004xx}
%  A.~Hebecker and A.~Westphal,
%``Gauge unification in extra dimensions: Power corrections vs.
%higher-dimension operators,''
Nucl.\ Phys.\ B {\bf 701}, 273 (2004)
[arXiv:hep-th/0407014].
%%CITATION = HEP-TH 0407014;%%

%
\bibitem{Arkani-Hamed:2001tb}
N.~Arkani-Hamed, T.~Gregoire and J.~Wacker,
%``Higher dimensional supersymmetry in 4D superspace,''
JHEP {\bf 0203} (2002) 055
[arXiv:hep-th/0101233].
%%CITATION = HEP-TH 0101233;%%

%
\bibitem{Marti:2001iw}
D.~Marti and A.~Pomarol,
%``Supersymmetric theories with compact extra dimensions in N = 1
%superfields,''
Phys.\ Rev.\ D {\bf 64} (2001) 105025
[arXiv:hep-th/0106256].
%%CITATION = HEP-TH 0106256;%%

%
\bibitem{arthur}
A.~Hebecker,
%``5D super Yang-Mills theory in 4-D superspace, superfield brane  operators,
%and applications to orbifold GUTs,''
Nucl.\ Phys.\ B {\bf 632} (2002) 101
[arXiv:hep-ph/0112230].
%%CITATION = HEP-PH 0112230;%%

%
%%CITATION = HEP-TH 0106256;%%
\bibitem{Kaplan:2001cg}
D.~E.~Kaplan and N.~Weiner,
%``Radion mediated supersymmetry breaking as a Scherk-Schwarz theory,''
arXiv:hep-ph/0108001;
%%CITATION = HEP-PH 0108001;%%
%\bibitem{vonGersdorff:2001ak}
G.~von Gersdorff and M.~Quiros,
%``Supersymmetry breaking on orbifolds from Wilson lines,''
Phys.\ Rev.\ D {\bf 65} (2002) 064016
[arXiv:hep-th/0110132].
%%CITATION = HEP-TH 0110132;%%

%
\bibitem{PaccettiCorreia:2004ri}
F.~Paccetti Correia, M.~G.~Schmidt and Z.~Tavartkiladze,
%``Superfield approach to 5D conformal SUGRA and the radion,''
Nucl.\ Phys.\ B {\bf 709} (2005) 141
[arXiv:hep-th/0408138];
%%CITATION = HEP-TH 0408138;%%
%``4D superfield reduction of 5D orbifold SUGRA and heterotic M-theory,''
arXiv:hep-th/0602173;
%%CITATION = HEP-TH 0602173;%%
%\bibitem{Abe:2005wn}
H.~Abe and Y.~Sakamura,
%``Scherk-Schwarz SUSY breaking from the viewpoint of 5D conformal
%supergravity,''
JHEP {\bf 0602} (2006) 014
[arXiv:hep-th/0512326].
%%CITATION = HEP-TH 0512326;%%

%
\bibitem{DGQ3}
D.~Diego, G.~von Gersdorff and M.~Quir\'os, work in progress.

%
\bibitem{vonGersdorff:2003rq}
G.~von Gersdorff, M.~Quiros and A.~Riotto,
%``Scherk-Schwarz supersymmetry breaking with radion stabilization,''
Nucl.\ Phys.\ B {\bf 689} (2004) 76
[arXiv:hep-th/0310190].
%%CITATION = HEP-TH 0310190;%%

%
\bibitem{Dimopoulos:1995mi}
S.~Dimopoulos and G.~F.~Giudice,
%``Naturalness constraints in supersymmetric theories with nonuniversal soft
%terms,''
Phys.\ Lett.\ B {\bf 357} (1995) 573
[arXiv:hep-ph/9507282].
%%CITATION = HEP-PH 9507282;%%

%
\bibitem{fine}
G.~W.~Anderson and D.~J.~Castano,
%``Measures of fine tuning,''
Phys.\ Lett.\ B {\bf 347} (1995) 300
[arXiv:hep-ph/9409419].
%%CITATION = HEP-PH 9409419;%%

%
%\cite{Wright:1998mk}
\bibitem{Wright:1998mk}
D.~Wright,
%``Naturally nonminimal supersymmetry,''
arXiv:hep-ph/9801449;
%%CITATION = HEP-PH 9801449;%%
%\bibitem{Kane:1998im}
G.~L.~Kane and S.~F.~King,
%``Naturalness implications of LEP results,''
Phys.\ Lett.\ B {\bf 451} (1999) 113
[arXiv:hep-ph/9810374].
%%CITATION = HEP-PH 9810374;%%

%
\bibitem{Eidelman:2004wy}
S.~Eidelman {\it et al.}  [Particle Data Group],
%``Review of particle physics,''
Phys.\ Lett.\ B {\bf 592} (2004) 1.
%%CITATION = PHLTA,B592,1;%%

%
\bibitem{Giudice:1995qk}
G.~F.~Giudice and A.~Pomarol,
%``Mass Degeneracy of the Higgsinos,''
Phys.\ Lett.\ B {\bf 372} (1996) 253
[arXiv:hep-ph/9512337].
%%CITATION = HEP-PH 9512337;%%

%
\bibitem{Gunion:1999jr}
J.~F.~Gunion and S.~Mrenna,
%``A study of SUSY signatures at the Tevatron in models with near mass
%degeneracy of the lightest chargino and neutralino,''
Phys.\ Rev.\ D {\bf 62} (2000) 015002
[arXiv:hep-ph/9906270].
%%CITATION = HEP-PH 9906270;%%
%\bibitem{Gunion:2001fu}
J.~F.~Gunion and S.~Mrenna,
%``Probing models with near degeneracy of the chargino and LSP at a linear  e+  %e- collider,''
Phys.\ Rev.\ D {\bf 64} (2001) 075002
[arXiv:hep-ph/0103167].
%%CITATION = HEP-PH 0103167;%%

%
\bibitem{Gondolo:2002tz}
P.~Gondolo, J.~Edsjo, P.~Ullio, L.~Bergstrom, M.~Schelke and E.~A.~Baltz,
%``DarkSUSY: A numerical package for supersymmetric dark matter
%calculations,''
arXiv:astro-ph/0211238.
%%CITATION = ASTRO-PH 0211238;%%

%
%\cite{Giudice:2004tc}
\bibitem{Giudice:2004tc}
G.~F.~Giudice and A.~Romanino,
%``Split supersymmetry,''
Nucl.\ Phys.\ B {\bf 699} (2004) 65
[Erratum-ibid.\ B {\bf 706} (2005) 65]
[arXiv:hep-ph/0406088].
%%CITATION = HEP-PH 0406088;%%

%
\bibitem{WMAP}
D.~N.~Spergel {\it et al.},
%``Wilkinson Microwave Anisotropy Probe (WMAP) three year results:
%Implications for cosmology,''
arXiv:astro-ph/0603449.
%%CITATION = ASTRO-PH 0603449;%%

%
\bibitem{diracity}
D.~R.~Smith and N.~Weiner,
%``Inelastic dark matter,''
Phys.\ Rev.\ D {\bf 64} (2001) 043502
[arXiv:hep-ph/0101138];
%%CITATION = HEP-PH 0101138;%%
%
I.~Antoniadis, A.~Delgado, K.~Benakli, M.~Quiros and M.~Tuckmantel,
%``Splitting extended supersymmetry,''
Phys.\ Lett.\ B {\bf 634} (2006) 302
[arXiv:hep-ph/0507192].
%%CITATION = HEP-PH 0507192;%%

%
%\cite{Weinberg:1982zq}
\bibitem{Weinberg:1982zq}
S.~Weinberg,
%``Cosmological Constraints On The Scale Of Supersymmetry Breaking,''
Phys.\ Rev.\ Lett.\  {\bf 48} (1982) 1303.
%%CITATION = PRLTA,48,1303;%%


\bibitem{gaugemediation}
  M.~Dine, W.~Fischler and M.~Srednicki,
  %``A Simple Solution To The Strong CP Problem With A Harmless Axion,''
  Phys.\ Lett.\ B {\bf 104} (1981) 199;
  %%CITATION = PHLTA,B104,199;%%
  M.~Dine and W.~Fischler,
  %``A Phenomenological Model Of Particle Physics Based On Supersymmetry,''
  Phys.\ Lett.\ B {\bf 110} (1982) 227;
  %%CITATION = PHLTA,B110,227;%%
  C.~R.~Nappi and B.~A.~Ovrut,
  %``Supersymmetric Extension Of The SU(3) X SU(2) X U(1) Model,''
  Phys.\ Lett.\ B {\bf 113} (1982) 175;
  %%CITATION = PHLTA,B113,175;%%
  L.~Alvarez-Gaume, M.~Claudson and M.~B.~Wise,
  %``Low-Energy Supersymmetry,''
  Nucl.\ Phys.\ B {\bf 207} (1982) 96;
  %%CITATION = NUPHA,B207,96;%%
  M.~Dine and A.~E.~Nelson,
  %``Dynamical supersymmetry breaking at low-energies,''
  Phys.\ Rev.\ D {\bf 48} (1993) 1277
  [arXiv:hep-ph/9303230];
  %%CITATION = HEP-PH 9303230;%%
  M.~Dine, A.~E.~Nelson and Y.~Shirman,
  %``Low-energy dynamical supersymmetry breaking simplified,''
  Phys.\ Rev.\ D {\bf 51} (1995) 1362
  [arXiv:hep-ph/9408384];
  %%CITATION = HEP-PH 9408384;%%
  M.~Dine, A.~E.~Nelson, Y.~Nir and Y.~Shirman,
  %``New tools for low-energy dynamical supersymmetry breaking,''
  Phys.\ Rev.\ D {\bf 53} (1996) 2658
  [arXiv:hep-ph/9507378].
  %%CITATION = HEP-PH 9507378;%%





\bibitem{Giudice:1998bp}
  G.~F.~Giudice and R.~Rattazzi,
  %``Theories with gauge-mediated supersymmetry breaking,''
  Phys.\ Rept.\  {\bf 322} (1999) 419
  [arXiv:hep-ph/9801271].
  %%CITATION = HEP-PH 9801271;%%

\end{thebibliography}
\end{document}